\documentclass[11pt]{article}
\usepackage{amsmath}
\usepackage{amssymb}
\usepackage{authblk}
\usepackage{bbm}
\usepackage{booktabs}
\usepackage[margin=1in]{geometry}
\usepackage{graphicx}
\usepackage[colorlinks=true]{hyperref}

\graphicspath{{figs/}}

\title{Multifidelity deep neural operators for efficient learning of partial differential equations with application to fast inverse design of nanoscale heat transport}
\author[1,*]{Lu Lu}
\author[2]{Rapha\"el Pestourie}
\author[2]{Steven G. Johnson}
\author[3]{Giuseppe Romano}
\affil[1]{Department of Chemical and Biomolecular Engineering, University of Pennsylvania, Philadelphia, PA 19104, USA}
\affil[2]{Department of Mathematics, Massachusetts Institute of Technology, Cambridge, MA 02139, USA}
\affil[3]{Institute for Soldier Nanotechnologies, Massachusetts Institute of Technology, Cambridge, MA 02139, USA}
\affil[*]{Corresponding author. Email: lulu1@seas.upenn.edu}
\date{}

\begin{document}

\maketitle

\begin{abstract}
Deep neural operators can learn operators mapping between infinite-dimensional function spaces via deep neural networks and have become an emerging paradigm of scientific machine learning. However, training neural operators usually requires a large amount of high-fidelity data, which is often difficult to obtain in real engineering problems. Here, we address this challenge by using multifidelity learning, i.e., learning from multifidelity datasets. We develop a multifidelity neural operator based on a deep operator network (DeepONet). A multifidelity DeepONet includes two standard DeepONets coupled by residual learning and input augmentation. Multifidelity DeepONet significantly reduces the required amount of high-fidelity data and achieves one order of magnitude smaller error when using the same amount of high-fidelity data. We apply a multifidelity DeepONet to learn the phonon Boltzmann transport equation (BTE), a framework to compute nanoscale heat transport. By combining a trained multifidelity DeepONet with genetic algorithm or topology optimization, we demonstrate a fast solver for the inverse design of BTE problems.
\end{abstract}

\section{Introduction}

Scientific machine learning (SciML) has become an emerging research field, where deep learning in the form of deep neural networks (DNNs) is used to address challenging problems in computational science and engineering~\cite{karniadakis2021physics}. In the past a few years, different SciML methods have been proposed~\cite{karniadakis2021physics}. One notable method is physics-informed neural networks (PINNs)~\cite{raissi2019physics,lu2021deepxde,karniadakis2021physics}, which have been developed to solve both forward and inverse problems of ordinary and partial differential equations (ODEs and PDEs). PINNs and the extensions~\cite{pang2019fpinns,zhang2019quantifying,lu2021physics,yu2021gradient} have been applied to solve diverse scientific and engineering applications~\cite{karniadakis2021physics,chen2020physics,yazdani2020systems,daneker2022systems}.

More recently, learning operators mapping between infinite-dimensional function spaces via DNNs has become a new and important paradigm of SciML~\cite{lu2019deeponet,lu2021learning,li2020fourier}. DNNs of this type are called deep neural operators. A typical application of deep neural operators for PDEs is used as a surrogate solver of the PDE, mapping from initial conditions, boundary conditions, or forcing terms to the PDE solution. Different architectures of deep neural operators have been developed, such as deep operator networks (DeepONet)~\cite{lu2019deeponet,lu2021learning,lu2021comprehensive}, Fourier neural operators~\cite{li2020fourier,lu2021comprehensive}, nonlocal kernel networks~\cite{you2022nonlocal}, and several others~\cite{trask2019gmls,li2020neural,patel2021physics}. Among these deep neural operators, DeepONet was the first one to be proposed (in 2019)~\cite{lu2019deeponet}, and many subsequent extensions and improvements have been developed, such as DeepONet with proper orthogonal decomposition (POD-DeepONet)~\cite{lu2021comprehensive}, DeepONet for multiple-input operators (MIONet)~\cite{jin2022mionet}, DeepONet for multi-physics problems via physics decomposition (DeepM\&Mnet)~\cite{cai2021deepm,mao2021deepm}, DeepONet with uncertainty quantification~\cite{lin2021accelerated,moya2022deeponet,yang2022scalable}, multiscale DeepONet~\cite{liu2021multiscale}, POD-DeepONet with causality~\cite{liu2022deeppropnet}, and physics-informed DeepONet~\cite{goswami2022physics,wang2021learning}.

DeepONet and its extensions have demonstrated good performance in diverse applications, such as fractional-derivative operators~\cite{lu2021learning}, stochastic differential equations~\cite{lu2021learning}, high-speed boundary-layer problems~\cite{di2021deeponet}, multiscale bubble growth dynamics~\cite{lin2021operator,lin2021seamless}, solar-thermal systems~\cite{osorio2022forecasting}, aortic dissection~\cite{yin2022simulating}, multiphysics and multiscale problems of electroconvection~\cite{cai2021deepm} and hypersonics~\cite{mao2021deepm}, power grids \cite{moya2022deeponet}, and multiscale modeling of mechanics problems~\cite{yin2022interfacing}. It has also been theoretically proved that DeepONet may break the curse of dimensionality in some PDEs~\cite{lanthaler2021error, deng2021convergence,marcati2021exponential}.

Despite this computational and theoretical progress, training DeepONet usually requires a large amount of data to achieve a good accuracy. However, in many engineering problems, it is often difficult to obtain necessary data of high accuracy. In these situations, it may be advantageous to complement the high-fidelity dataset (small and accurate) by employing a low-fidelity dataset (larger and less accurate), i.e., learning from multifidelity datasets~\cite{fernandez2016review,meng2020composite,lu2020extraction}. Multifidelity learning can be used in diverse applications and scenarios. For example, the high-fidelity dataset might be obtained from simulations with fine mesh, while the low-fidelity dataset is from simulations with coarse mesh. In Ref.~\cite{lu2020extraction}, the high-fidelity dataset is from experiments and 3D simulations, while the low-fidelity dataset is from 2D simulations and analytical solutions. Multifidelity learning by neural networks has been developed for function approximation in Refs.~\cite{meng2020composite,lu2020extraction}. Another recent approach to data efficiency using a low-fidelity solver trains a combination of a generative neural network and a low-fidelity solver, end-to-end, to match the limited data of a high-fidelity solver~\cite{pestourie2021physics}.

In this work, we aim to alleviate the challenge of large datasets required for training DeepONet. We propose new algorithms to significantly reduce the required number of high-fidelity data. Specifically, we develop a new version of DeepONet by using residual learning and input augmentation, that is capable of fusing together different sets of data with different fidelity levels, making the proposed multifidelity DeepONet very efficient for learning complex systems.

As an application, we apply multifidelity DeepONet to learn heat transport in nanostructured materials, described by the phonon Boltzmann-transport equation (BTE)~\cite{ziman2001electrons}. We first train a multifidelity DeepONet to predict the map of magnitude of thermal flux in a Si nanoporous system; then, the trained DeepONet is employed with genetic algorithms and topology optimization for inverse design of materials to maximize heat transport over prescribed points with a constrained number of pores. Our method is validated against numerical BTE simulations. We note that the same DeepONet can be used for multiple inverse design tasks without the need for retraining. Potential applications in this area include energy harvesting and heat management~\cite{li2012colloquium,narayana2012heat,kadic2013metamaterials}.

The paper is organized as follows. In Section~\ref{sec:method}, after introducing the algorithm of DeepONet, we present the setup of multifidelity learning and a few different approaches of multifidelity DeepONet. Then we discuss how to apply the trained DeepONet for inverse design. In Section~\ref{sec:results}, we compare different approaches of multifidelity DeepONet and demonstrate the effectiveness on a Poisson equation and BTE problem. Subsequently, we solve several inverse design problems of BTE for different objective functions. Finally, we conclude the paper in Section~\ref{sec:conclusion}.

\section{Methods}
\label{sec:method}

We first introduce DeepONet and then present different approaches of multifidelity learning for DeepONet. We also discuss how to use the learned multifidelity DeepONet for inverse design.

\subsection{DeepONet}

DeepONet was proposed in Ref.~\cite{lu2021learning} for learning nonlinear operators mapping from a function to another function, e.g., from the boundary/initial condition to the PDE solution. We denote the input function by $v$ defined on the domain $D \in \mathbb{R}^d$:
$$v: D \ni x \mapsto v(x) \in \mathbb{R},$$
and the output function by $u$ defined on the domain $D' \in \mathbb{R}^{d'}$:
$$u: D' \ni \xi \mapsto u(\xi) \in \mathbb{R}.$$
Let $\mathcal{V}$ and $\mathcal{U}$ be the spaces of $v$ and $u$, respectively, and then the operator is
$$\mathcal{G}: \mathcal{V} \ni v \mapsto u \in \mathcal{U}.$$

\begin{figure}[htbp]
    \centering
    \includegraphics{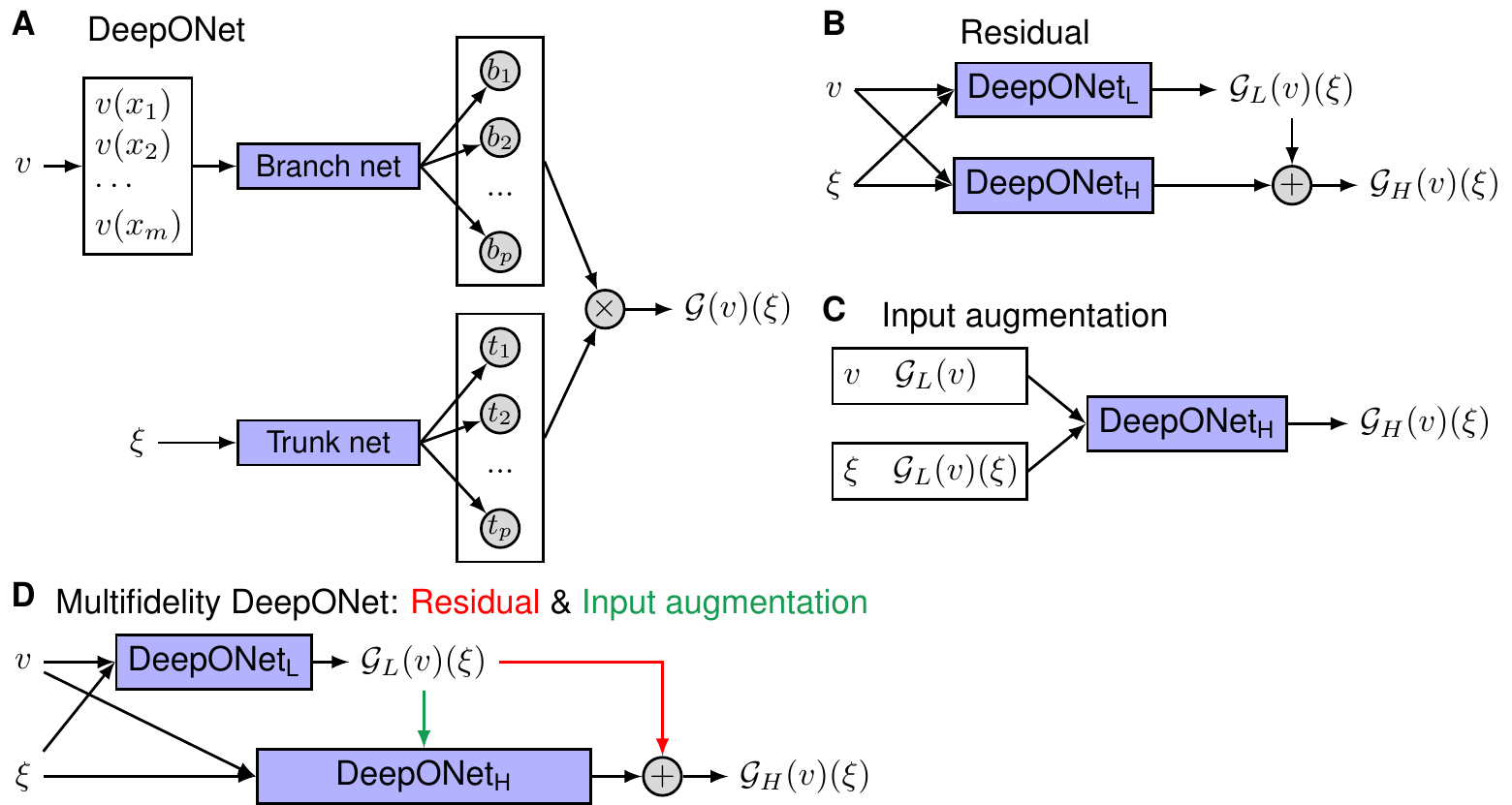}
    \caption{\textbf{Architectures of DeepONet and multifidelity DeepONet.} (\textbf{A}) Architecture of DeepONet, adapted from~\cite{lu2021learning}. (\textbf{B}) A DeepONet learns the residual between the high-fidelity and low-fidelity solutions. (\textbf{C}) The low-fidelity solution is used as the extra input of the DeepONet. (\textbf{D}) Multifidelity DeepONet uses the residual learning (the red line) and input augmentation (the green line).}
    \label{fig:deeponet}
\end{figure}

DeepONet is used to learn the mapping $\mathcal{G}$ from a dataset of $v$ and $u$. In a DeepONet, there are two sub-networks: a branch net and a trunk net (Fig.~\ref{fig:deeponet}A). The branch net takes the evaluations of $v$ at $\{x_1, x_2, \dots, x_m\}$ as its input, while the trunk net takes $\xi$ as its input. Then the DeepONet output is an inner product of the branch and trunk outputs
$$\mathcal{G}(v)(\xi) = \sum_{k=1}^p b_k(v) t_k(\xi) + b_0,$$
where $b_0$ as a bias.

\subsection{Multifidelity DeepONet}
\label{sec:mf_deeponet}

DeepONet can be used to learn diverse nonlinear operators, but it usually requires a large high-fidelity dataset for training, which is generated by expensive numerical simulations. Here, we propose multifidelity DeepONet to reduce dramatically the required high-fidelity dataset by leveraging an additional low-fidelity dataset.

\subsubsection{Multifidelity learning}

The idea of multifidelity learning is that instead of generating a large dataset of high accuracy (i.e., high-fidelity), we only generate a small high-fidelity dataset, but in the meanwhile, we generate another dataset of low accuracy (i.e., low-fidelity). The low-fidelity dataset is much cheaper to generate, and thus it is easy to generate a large low-fidelity dataset. In short, in multifidelity learning, we have two datasets:
\begin{itemize}
    \item a high-fidelity dataset of size $N_H$:
    $$\mathcal{T}_H = \left\{ \left(v^H_i, u^H_i = \mathcal{G}_H\left(v^H_i\right)\right) \right\}_{i=1, 2, \dots, N_H}$$
    \item a low-fidelity dataset of size $N_L$:
    $$\mathcal{T}_L =\left\{ \left(v^L_i, u^L_i = \mathcal{G}_L\left(v^L_i\right)\right) \right\}_{i=1, 2, \dots, N_L}$$
\end{itemize}
where $N_H \ll N_L$. $\mathcal{G}_H$ (i.e., the original $\mathcal{G}$) is the high-fidelity operator we aim to learn, and $\mathcal{G}_L$ is the low-fidelity operator. Then we use both datasets to learn the operator.

Because we have a large low-fidelity dataset, it is straightforward to use a DeepONet (DeepONet\textsubscript{L}) to learn $\mathcal{G}_L$ from $\mathcal{T}_L$. Here, we develop two methods to learn $\mathcal{G}_H$, including residual learning and input augmentation.

\subsubsection{Residual learning}
\label{sec:residual}

Although the low-fidelity solution is not accurate enough, it captures the basic trend or shape of the solution. Instead of directly learning the high-fidelity output, we learn the residual between the high-fidelity and low-fidelity outputs:
$$\mathcal{R}(v)(\xi) = \mathcal{G}_H(v)(\xi) - \mathcal{G}_L(v)(\xi).$$
We use another DeepONet (DeepONet\textsubscript{H}) to learn the residual operator $\mathcal{R}$. Hence, we have two DeepONets (DeepONet\textsubscript{L} and DeepONet\textsubscript{H}) (Fig.~\ref{fig:deeponet}B). DeepONet\textsubscript{L} is trained using $\mathcal{T}_L$, and DeepONet\textsubscript{H} is trained using $\mathcal{T}_H$. During the preparation of this manuscript, a new paper \cite{de2022bi} appeared, which also proposed a similar approach of residual learning for multifidelity DeepONets.

\subsubsection{Input augmentation}
\label{sec:input_aug}

For a same input function $v$, the low-fidelity solution $\mathcal{G}_L(v)$ and high-fidelity solution $\mathcal{G}_H(v)$ are usually highly correlated. In our residual learning above, therefore, we are exploiting the property that $\mathcal{G}_H(v)$ and $\mathcal{G}_L(v)$ exhibit similar behaviors, which should make their difference $\mathcal{R}$ is easy to learn. We can also learn this correlation directly from data. Specifically, we use the low-fidelity prediction as an additional input of DeepONet\textsubscript{H}. We can append the low-fidelity prediction to the input of the branch net and/or the trunk net (Fig.~\ref{fig:deeponet}C).

\paragraph{Approach I.} We append the low-fidelity prediction $\mathcal{G}_L(v)$ (the entire function) to the branch net inputs:
$$\mathcal{G}_H(v)(\xi) = \mathcal{Q}(v, \mathcal{G}_L(v))(\xi),$$
i.e., the branch net has two functions $v$ and $\mathcal{G}_L(v)$ as the input. In this study, we directly concatenate the discretized $v$ and $\mathcal{G}_L(v)$.

\paragraph{Approach II.} Instead of using the entire function $\mathcal{G}_L(v)$, we can also use only the evaluation of $\mathcal{G}_L(v)$ at the location $\xi$ to be predicted, i.e., $\mathcal{G}_L(v)(\xi)$. In this case, we append $\mathcal{G}_L(v)(\xi)$ in the trunk-net inputs:
$$\mathcal{G}_H(v)(\xi) = \mathcal{Q}(v)(\xi, \mathcal{G}_L(v)(\xi)).$$

In our numerical experiments with the Poisson equation in Section~\ref{sec:poisson}, we find that Approach~II achieves better accuracy than Approach~I.

\subsubsection{Residual learning and input augmentation}

We also combine the methods of residual learning and input augmentation Approach~II (Fig.~\ref{fig:deeponet}D), where we have two DeepONets: DeepONet\textsubscript{L} and DeepONet\textsubscript{H}. DeepONet\textsubscript{L} learns the low-fidelity operator $\mathcal{G}_L$, and DeepONet\textsubscript{H} learns the residual operator $\mathcal{R}$. DeepONet\textsubscript{H} also uses $\mathcal{G}_L(v)(\xi)$ as one of its inputs of the trunk net. The final prediction of the high-fidelity solution is
\begin{equation} \label{eq:Ghigh}
\mathcal{G}_H(v)(\xi) = \underbrace{\mathcal{G}_L(v)(\xi)}_{\text{DeepONet\textsubscript{L}}} + \underbrace{\mathcal{R}(v)(\xi, \overbrace{\mathcal{G}_L(v)(\xi)}^{\text{DeepONet\textsubscript{L}}})}_{\text{DeepONet\textsubscript{H}}}.  
\end{equation}

We can decompose the error of the multifidelity DeepONet $\mathcal{E}_{\mathcal{G}_H}$ by the triangle inequality as
\begin{equation} \label{eq:error}
    \mathcal{E}_{\mathcal{G}_H} \le \mathcal{E}_{\mathcal{G}_L} + \mathcal{E}_{\mathcal{R}},
\end{equation}
where $\mathcal{E}_{\mathcal{G}_L}$ and $\mathcal{E}_{\mathcal{R}}$ are the errors of DeepONet\textsubscript{L} and DeepONet\textsubscript{H}, respectively. Hence, in order to have an accurate multifidelity DeepONet surrogate for $\mathcal{G}_H$, both DeepONet\textsubscript{L} and DeepONet\textsubscript{H} should be trained well.

\subsection{Inverse design}
\label{sec:inverse}

Once we have a surrogate model of the operator $\mathcal{G}$, we can predict $\mathcal{G}(v)$ for any $v$ within a fraction of second. One application is inverse design, which is formulated as an optimization problem:
$$\max_{v} \mathcal{J}(u; v),$$
subject to
$$u = \mathcal{G}(v).$$
We have the flexibility to choose the optimization algorithm.

\subsubsection{Genetic algorithm}

We can use derivative-free optimization algorithm such as genetic algorithm (GA). In this study, we consider the canonical genetic algorithm. The algorithm starts from a population of randomly generated candidate solutions (called individuals), and then the population is evolved toward better solutions. The population in each iteration is called a generation. In each generation, the individuals are selected and modified via the operations of mutation and crossover to form a new generation in the next iteration. The individuals with higher values of the objective function have higher chance to be selected. We terminate the algorithm until a maximum number of generations has been produced. For more details of the algorithm, see Ref.~\cite{back2018evolutionary}. We implement the algorithm using the library DEAP~\cite{DEAP_JMLR2012}. In this work, we choose the population size as 100.

\subsubsection{Gradient-based topology optimization}

Because $\mathcal{G}$ is approximated by DeepONets in Eq.~\eqref{eq:Ghigh}, we can compute the derivative of $\frac{\partial u(\xi)}{\partial v(x_i)}$ as
$$\frac{\partial \mathcal{G}_H(v)(\xi)}{\partial v(x_i)} = \frac{\partial \mathcal{R}(v)(\xi, \mathcal{G}_L(v)(\xi))}{\partial v(x_i)} + \frac{\partial \mathcal{G}_L(v)(\xi)}{\partial v(x_i)}\left( 1+\frac{\partial \mathcal{R}(v)(\xi, \mathcal{G}_L(v)(\xi))}{\partial\mathcal{G}_L(v)(\xi)} \right),$$
where the three partial derivatives in the right-hand side are computed by automatic differentiation (AD; also called ``backpropagation''). Then we can compute $\frac{\partial \mathcal{J}}{\partial v(x_i)}$ by the chain rule and use gradient-based topology optimization (TO) methods.

In TO, each pixel of the domain is a degree of freedom, where the material property is iteratively updated using a gradient-based optimization algorithm, such as the conservative convex separable approximations algorithm~\cite{svanberg2002class}.
In contrast to genetic algorithm, TO uses a continuous relaxation optimization scheme, which takes full advantage of the extension from binary inputs to real inputs of neural networks. Even though the data only has binary inputs, the neural network defines a differentiable continuous function which can return function evaluations and gradient evaluations anywhere between the binary inputs.
Despite the fact that this continuation is artificial, i.e., non-binary values do not correspond to any physical materials, we can use this framework to perform optimization as long as the end design only contains binary values.

In order to converge to a binary-input result, TO uses a smoothed approximation of the Heaviside function as a thresholding function, which pushes the design values towards binary values~\cite{christiansen2021inverse}:
\begin{equation} \label{eq:thresholding}
    \tilde{\xi}(\xi)=\frac{\tanh{\beta \eta}+\tanh{\beta(\xi-\eta)}}{\tanh{\beta \eta}+\tanh{\beta(1-\eta)}},
\end{equation}
where $\eta$ is a hyperparameter to be fixed in $[0,1]$, and $\beta\in [1, \infty]$ controls how binary the resulting design will be. 
The larger the $\beta$, the stiffer the optimization becomes. In TO, $\beta$ starts at 1, and is successively doubled after each round of optimization, using the optimal design of the last round of optimization as a starting point to the next, until the optimum converges to binary values. Although the thresholding function is usually sufficient to converge to binary design, it is not guaranteed. In the case when the thresholding function is not sufficient to converge to a binary design, a penalty function $\bar{\xi}(\tilde{\xi}) = \tilde{\xi} (1-\tilde{\xi})$  is added to the objective function to force the design to be binary.
% In our numerical experiments, we show that TO successfully finds the same optimum as genetic algorithm for small problem, but TO can also scale to problem with orders of magnitude more inputs via large-scale gradient-based optimization, while genetic algorithm suffers the ``curse of dimensionality''.

\section{Results}
\label{sec:results}

We apply the proposed multifidelity DeepONet to learn the Poisson equation and Boltzmann transport equation. We then use the learned multifidelity DeepONet for inverse design of thermal transport in nanostructures. All the multifidelity DeepONet codes in this study are implemented by using the library DeepXDE~\cite{lu2021deepxde} and will be deposited in GitHub at \url{https://github.com/lu-group/multifidelity-deeponet}.

% \subsection{Learning the antiderivative operator}

% First, we consider a pedagogical example described by
% $$\frac{du(x)}{dx} = f(x)$$
% with an initial condition $u(0)=0$. We learn the operator mapping from $f(x)$ to $u(x)$, i.e.,
% $$\mathcal{G}: f(x) \mapsto u(y) = \int_0^y f(x)dx.$$

% The low-fidelity dataset is obtained by solving the ODE using the forward Euler method with a step size $h_{Euler}=0.2$, and the high-fidelity dataset is computed using the Runge-Kutta method of order 5(4) with a step size $h_{RK}=0.05$.

\subsection{Poisson equation}
\label{sec:poisson}

We first consider a 1D Poisson equation to compare the different multifidelity methods proposed in Section~\ref{sec:mf_deeponet}, and demonstrate the effectiveness of multifidelity DeepONet.

\subsubsection{Problem setup}

\begin{figure}[htbp]
    \centering
    \includegraphics{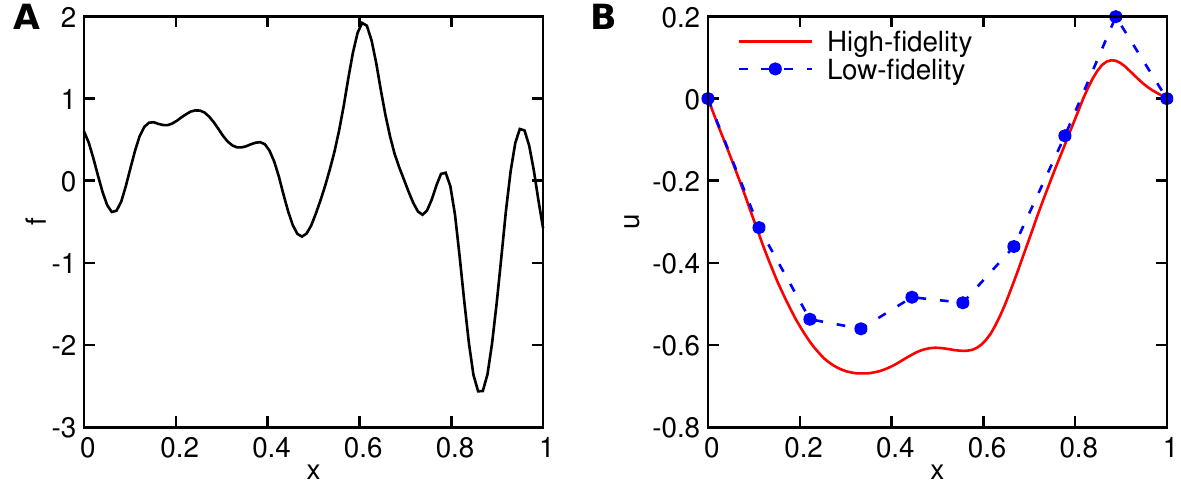}
    \caption{\textbf{Examples of the Poisson equation.} (\textbf{A}) An example of randomly sampled $f$. (\textbf{B}) The corresponding low- and high-fidelity solutions.}
    \label{fig:poisson_example}
\end{figure}

The 1D Poisson equation is described by
$$\frac{d^2 u}{dx^2} = 20 f(x), \quad x \in [0, 1],$$
with the Dirichlet boundary condition $u(0) = u(1) = 0$. We aim to learn the operator from the forcing term $f(x)$ to the PDE solution:
$$\mathcal{G}: f \mapsto u.$$
We sample $f$ from a Gaussian random field (GRF) with mean zero:
$$f \sim \mathcal{GP}(0, k_l(x_1, x_2)),$$
where the covariance kernel $k_l(x_1, x_2) = \exp\left(-|x_1-x_2|^2/(2l^2)\right)$ is the Gaussian kernel with a length-scale parameter $l=0.05$. An example of randomly sampled $f$ is shown in Fig.~\ref{fig:poisson_example}A.

We generate the high-fidelity and low-fidelity datasets by solving the Poisson equation via the finite difference method with different mesh size $\Delta x$. For the high-fidelity solutions, we use $\Delta x = 1/99$, and for the low-fidelity solutions, $\Delta x = 1/9$. The low- and high-fidelity solutions for the randomly sample $f$ above are shown in Fig.~\ref{fig:poisson_example}B, which shows that the low-fidelity solution has a large error. When testing on a large dataset, the mean squared error (m.s.e.) of the low-fidelity solver is 0.0233.

\subsubsection{High-fidelity only}

Instead of achieving high accuracy, we aim to compare the performance of different approaches of multifidelty DeepONet, and thus we generate an extremely small high-fidelity dataset. In the high-fidelity dataset, we have 500 different samples of $f$, but for each $f$, we do not have the full field observation of the corresponding solution $u$, and instead, we only know the value of $u(x)$ at one location $x$ randomly sampled in $[0, 1]$.

\begin{figure}[htbp]
    \centering
    \includegraphics{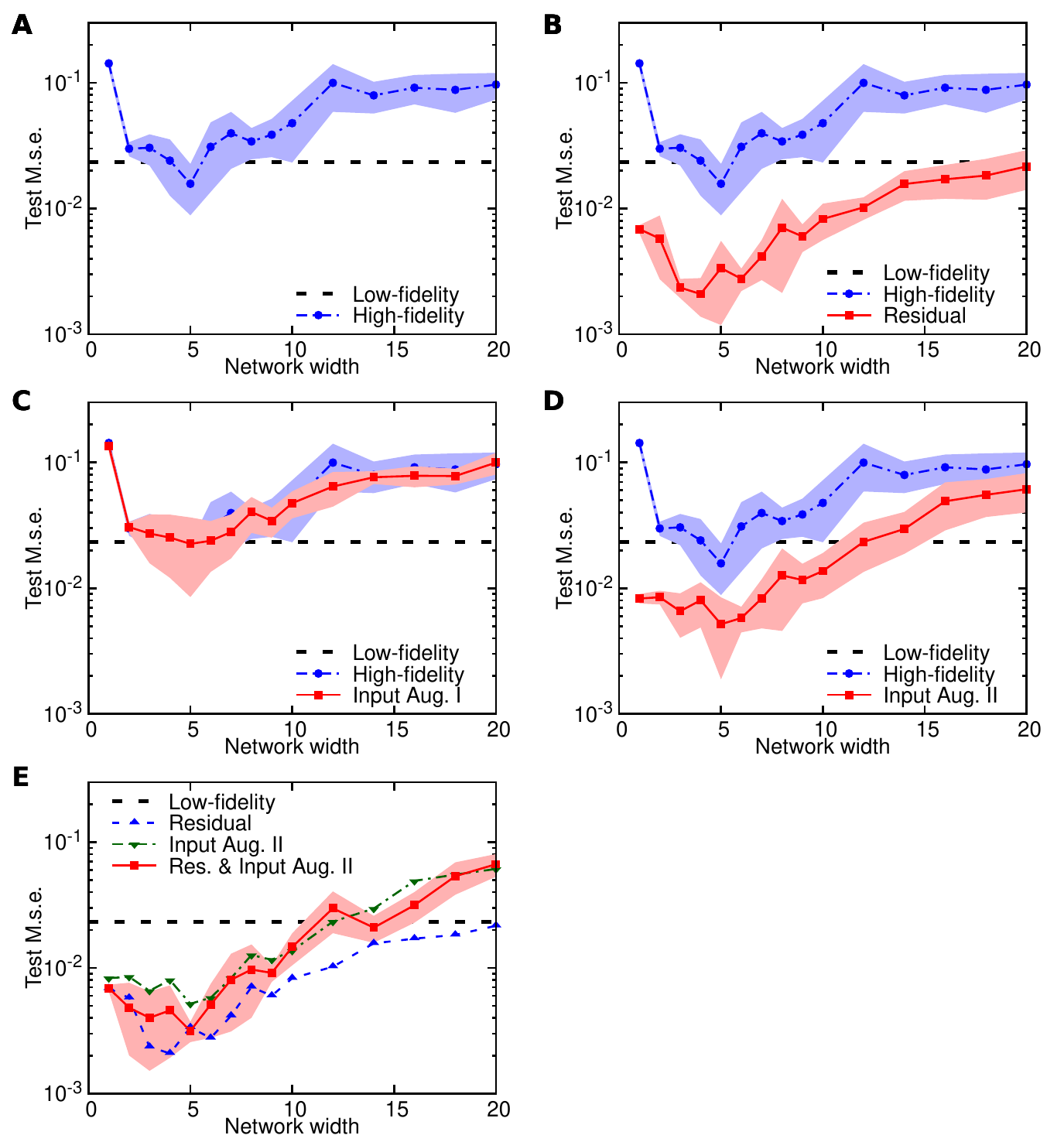}
    \caption{\textbf{Comparison of different multifidelity methods for DeepONet on the Poisson equation.} (\textbf{A}) DeepONet trained only with an extremely small high-fidelity dataset. (\textbf{B}) Multifidelity DeepONet with residual learning. (\textbf{C}) Multifidelity DeepONet with input augmentation Approach I. (\textbf{D}) Multifidelity DeepONet with input augmentation Approach II. (\textbf{E}) Multifidelity DeepONet with residual learning and input augmentation Approach II. The shaded regions represent the one standard deviation of 10 runs with randomly generated dataset and random network initialization.}
    \label{fig:poisson}
\end{figure}

Because the dataset is small, we choose both the trunk net and the branch net as shallow fully-connected neural networks with the SELU activation function~\cite{klambauer2017self}. All DeepONets are trained with Adam optimizer~\cite{kingma2014adam} with a learning rate $10^{-4}$ for 50000 epochs. We test DeepONets with different width, and the DeepONet with the width 5 has the smallest test m.s.e. of 0.0157 $\pm$ 0.0069 (Fig.~\ref{fig:poisson}A). The mean and standard deviation of the error are obtained by 10 runs with randomly generated dataset and random network initialization. Larger DeepONets have the issue of over-fitting. Because an extremely small dataset has been used, the best DeepONet has similar accuracy as the low-fidelity solution.

\subsubsection{Multifidelity}
\label{sec:poisson_mf}

We use the same set up (the same high-fidelity dataset and the same hyperparameters) to test different approaches of multifidelity DeepONet. Because a multifidelity DeepONet has a low-fidelity DeepONet (DeepONet\textsubscript{L}) and a high-fidelity DeepONet (DeepONet\textsubscript{H}), in order to remove the interference of DeepONet\textsubscript{L}, we directly use the low-fidelity solver to replace DeepONet\textsubscript{L} in Figs.~\ref{fig:deeponet}B, C and D, i.e., we assume that the network DeepONet\textsubscript{L} is perfect. Because the low-fidelity solver can only predict the solution on the 10 locations as shown in Fig.~\ref{fig:poisson_example}B, we perform a linear interpolation to compute the low-fidelity solution for other locations (e.g., the dash line in Fig.~\ref{fig:poisson_example}B).

\paragraph{Residual learning.} We first test the residual learning in Section~\ref{sec:residual}. By learning the residual, the error of DeepONet is always about one order of magnitude smaller than the DeepONet with high-fidelity only (Fig.~\ref{fig:poisson}B) no matter what the network width is. The smallest error (0.0021 $\pm$ 0.0007) is obtained for the width 4.

\paragraph{Input augmentation.} Similarly, we also test the two approaches of input augmentation in Section~\ref{sec:input_aug}. We show that the approach I, i.e., using the entire low-fidelity solution as the branch net input, does not have any improvement (Fig.~\ref{fig:poisson}C). For the approach II of only using the low-fidelity solution of the target point as the trunk net input, the smallest test error is 0.0052 $\pm$ 0.0033, also obtained at the width 5. Input augmentation approach II is uniformly better than high-fidelity only (Fig.~\ref{fig:poisson}D).

\paragraph{Residual and Input augmentation (Approach I).} We further combine the residual learning and the input augmentation (Approach II). When the width is 5, the smallest test m.s.e. is 0.0031 $\pm$ 0.0006 (Fig.~\ref{fig:poisson}E).

We show that in this Poisson equation, residual learning and input augmentation approach II improve the accuracy of DeepONet, but there is no improvement by combing them. However, as we show in the next section, by using both approaches together, we can achieve better accuracy.

\subsection{Boltzmann transport equation}

When the characteristic length of a semiconductor material approaches the mean-free-path (MFP) of heat carriers, i.e., phonons, Fourier's law breaks down~\cite{ziman2001electrons}. Here, we learn the nondiffusive heat transport with the steady-state phonon Boltzmann transport equation (BTE) by multifidelity DeepONet.

\subsubsection{Problem setup}

Under the relaxation time approximation, BTE is given by~\cite{ziman2001electrons}
\begin{equation} \label{eq:bte}
    -\mathbf{v}_\mu \cdot\nabla f_\mu(\mathbf{r}) = \frac{f_\mu(\mathbf{r}) - f^0_\mu(\mathbf{r})}{\tau_\mu},
\end{equation}
where $\mathbf{v}_\mu$ and $\tau_\mu$ are the phonon group velocity and intrinsic scattering time, respectively. The label $\mu$ indicates both the phonon wave vector and polarization. The unknown of Eq.~\eqref{eq:bte} is the nonequilibrium phonon distribution $f_\mu(\mathbf{r})$. Both the group velocities and the scattering times, computed with density functional theory (see Ref.~\cite{lindsay2019perspective} for its application to phonon transport), are obtained by AlmaBTE~\cite{carrete2017almabte} using a wave-vector discretization of $24 \times 24 \times 24$. The equilibrium distribution, $f^0_\mu(\mathbf{r})$, under the assumption of small temperature variation, is given by $$f^0_\mu(\mathbf{r})=\sum_{\mu'}\alpha_{\mu'}f_{\mu'}(\mathbf{r}),$$
where $\alpha_{\mu'} = C_{\mu'}/\tau_{\mu'}\left(\sum_{\mu''}C_{\mu''}/\tau_{\mu''} \right)^{-1}$; the terms $C_\mu$ and $\tau_\mu$ are the mode-resolved heat capacity and scattering time, respectively~\cite{carrete2017almabte}. 

\begin{figure}[htbp]
    \centering
    \includegraphics{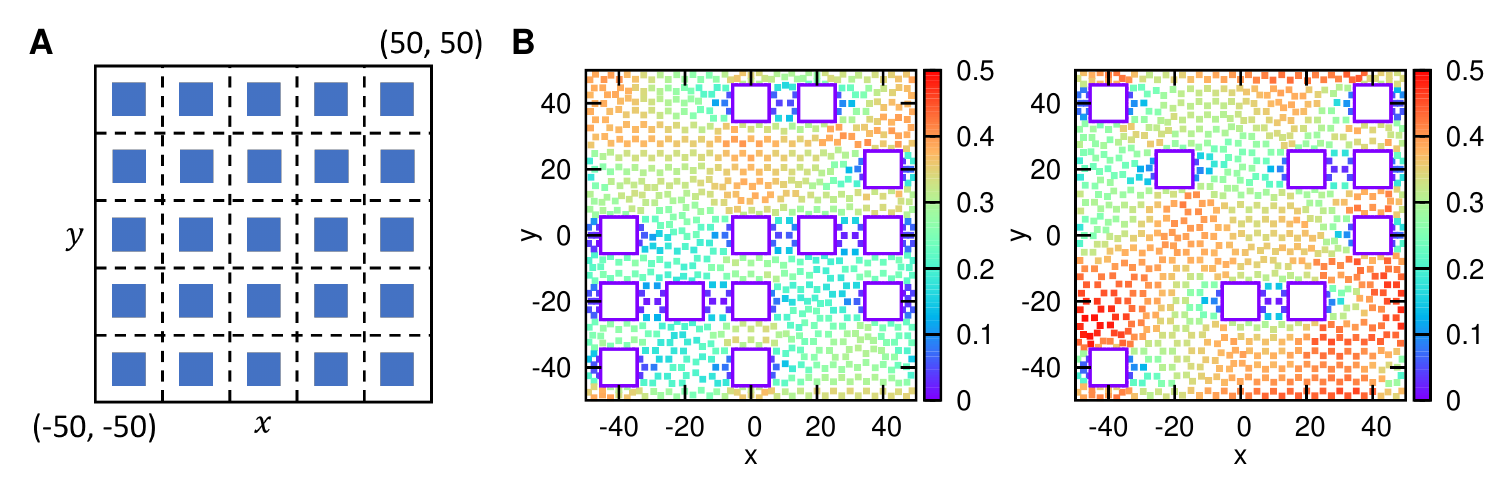}
    \caption{\textbf{BTE problem setup and two examples.} (\textbf{A}) The computational domain and possible pore locations. (\textbf{B}) Two examples of the pore locations and the corresponding flux. The dots represent the locations of the mesh nodes.}
    \label{fig:bte}
\end{figure}

We consider BTE defined on a rectangular domain (in nm) $\Omega = [-50, 50] \times [-50, 50]$ with periodic boundary conditions in both $x$ and $y$ directions, and a temperature difference of 1K is applied along the \textit{x}-axis. Inside the domain, we have a five by five equispaced arrays of small squares (the blue squares in Fig.~\ref{fig:bte}A). Each square is of size (10, 10) and can be a pore without any material. Along the wall of the pores, we adopt the totally diffuse boundary conditions~\cite{romano2021efficient}.

We solve Eq.~\eqref{eq:bte} by the finite-volume method implemented in the free/open-source software OpenBTE~\cite{romano2021openbte}, which adopts the source iteration scheme~\cite{romano2021efficient}, i.e.,
\begin{equation} \label{eq:si}
    \tau_\mu \mathbf{\mathbf{v}_\mu}\cdot \nabla f_\mu^{(n)}(\mathbf{r}) +f_\mu^{(n)}(\mathbf{r}) = \sum_{\mu'} a_{\mu'}f_{\mu'}^{(n-1)}(\mathbf{r}).
\end{equation}
Here, we are interested in the heat flux $\mathbf{J} = \frac{1}{\mathcal{V}}\sum_\mu C_\mu \mathbf{v}_\mu f_\mu $ ($\mathcal{V}$ being a normalization volume), and aim to learn the flux for different locations of pores. We generate the dataset by randomly sampling the locations of pores. For different number and locations of the pores, we have different BTE solutions. Each data point in the dataset is a pair of the pore locations and the corresponding solution of flux in the mesh nodes. There are in total $2^{25} \approx 3.4\times 10^7$ possibilities, and two examples of the random pore locations and the corresponding flux is shown in Fig.~\ref{fig:bte}B. Different geometry has different mesh, and the mesh nodes of the two examples are the dots in Fig.~\ref{fig:bte}B. In average, each mesh has 902 nodes. The high-fidelity solutions are obtained by Eq.~\eqref{eq:si} solved for 5 iterations, while the low-fidelity solutions are obtained with only 2 iterations.

In DeepONet, the branch net input $v = (v_{1,1}, v_{1,2}, \dots, v_{5,5})$ represents the locations of pores. If $v_{i,j}=1$, then it is a pore, otherwise it is not a pore. We note that when training DeepONet, we scale the $x$ and $y$ coordinates from $[-50, 50]$ to $[-1, 1]$ and also normalize the network output to zero mean and unit variance.

\subsubsection{Effectiveness of multifidelity learning}

\begin{figure}[htbp]
    \centering
    \includegraphics{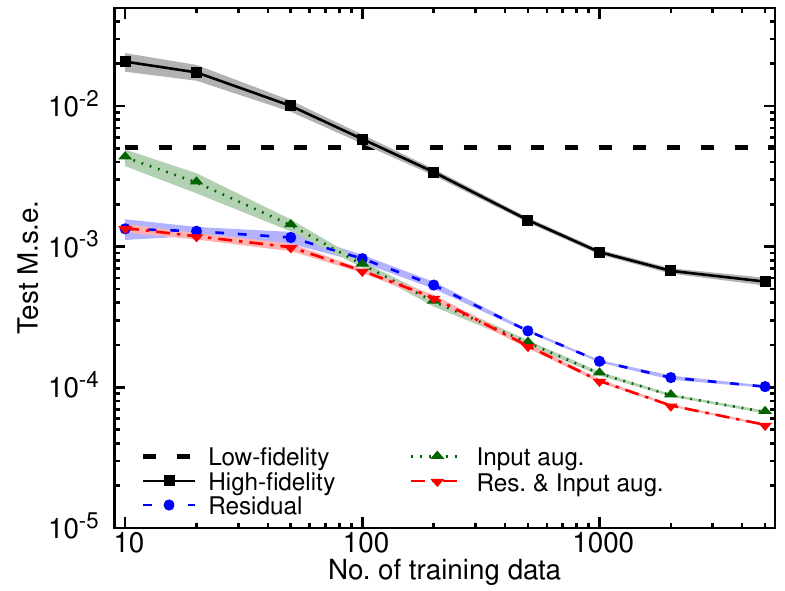}
    \caption{\textbf{Comparison of different multifidelity methods for DeepONet on BTE.}}
    \label{fig:bte_mf}
\end{figure}

We use both the low-fidelity solver and DeepONet only trained with high-fidelity data as the two baselines. Compared to the high-fidelity solution, the low-fidelity solution has m.s.e. of $5.09\times 10^{-3}$ (the horizontal dash line in Fig.~\ref{fig:bte_mf}). We train a DeepONet with only high-fidelity solutions, where the branch net and the trunk net use ReLU activation function and have 4 layers and 5 layers, respectively. DeepONet is trained with Adam optimizer~\cite{kingma2014adam} with a learning rate $10^{-4}$ for 100000 epochs. The DeepONet needs about 150 training data to achieve similar accuracy of the low-fidelity solver (the solid black line in Fig.~\ref{fig:bte_mf}).

We then compare different approaches of multifidelity DeepONet. As discussed in Section~\ref{sec:poisson_mf}, for better comparison, here we still use the low-fidelity solver to replace DeepONet\textsubscript{L}. In the same setup, both residual learning and input augmentation approach II improve the accuracy by one order of magnitude (Fig.~\ref{fig:bte_mf}). When the dataset is very small ($< 100$), residual learning is better than input augmentation, otherwise input augmentation becomes better. When we use both approaches together, we have even better accuracy.

\subsubsection{Multifidelity DeepONet}
\label{sec:bte_mf_deeponet}

We have demonstrated the effectiveness of multifidelity DeepONet. As we mentioned above, we directly used the exact low-fidelity solver, and here we first train a low-fidelity DeepONet (DeepONet\textsubscript{L}) and then combine it with DeepONet\textsubscript{H} to have the multifidelity DeepONet.

\paragraph{Low-fidelity DeepONet.} DeepONet\textsubscript{L} is trained with a dataset of size 10000 solutions. We perform a hyperparameter tuning and choose the branch net as 3 layers and the trunk net as 5 layers. Both the branch net and the trunk net use ReLU activation function and have width 512. After training with Adam optimizer~\cite{kingma2014adam} with a learning rate $10^{-4}$ for 500000 epochs, the testing m.s.e. is $4.61 \pm 0.21 \times 10^{-5}$, and the $L^2$ relative error is 1.91$\pm$0.02\%.

The flux satisfies the periodic boundary conditions in $x$ and $y$ directions, and as proposed in Ref.~\cite{lu2021comprehensive}, we can enforce the periodic BCs in DeepONet by adding a Fourier feature layer. Specifically, we use the basis functions of the Fourier series on a 2D square as the first layer of the trunk net. By enforcing PBC, test m.s.e. decreases to $3.90 \pm 0.15 \times 10^{-5}$, and $L^2$ relative error becomes 1.72$\pm$0.02\%.

\begin{figure}[htbp]
    \centering
    \includegraphics{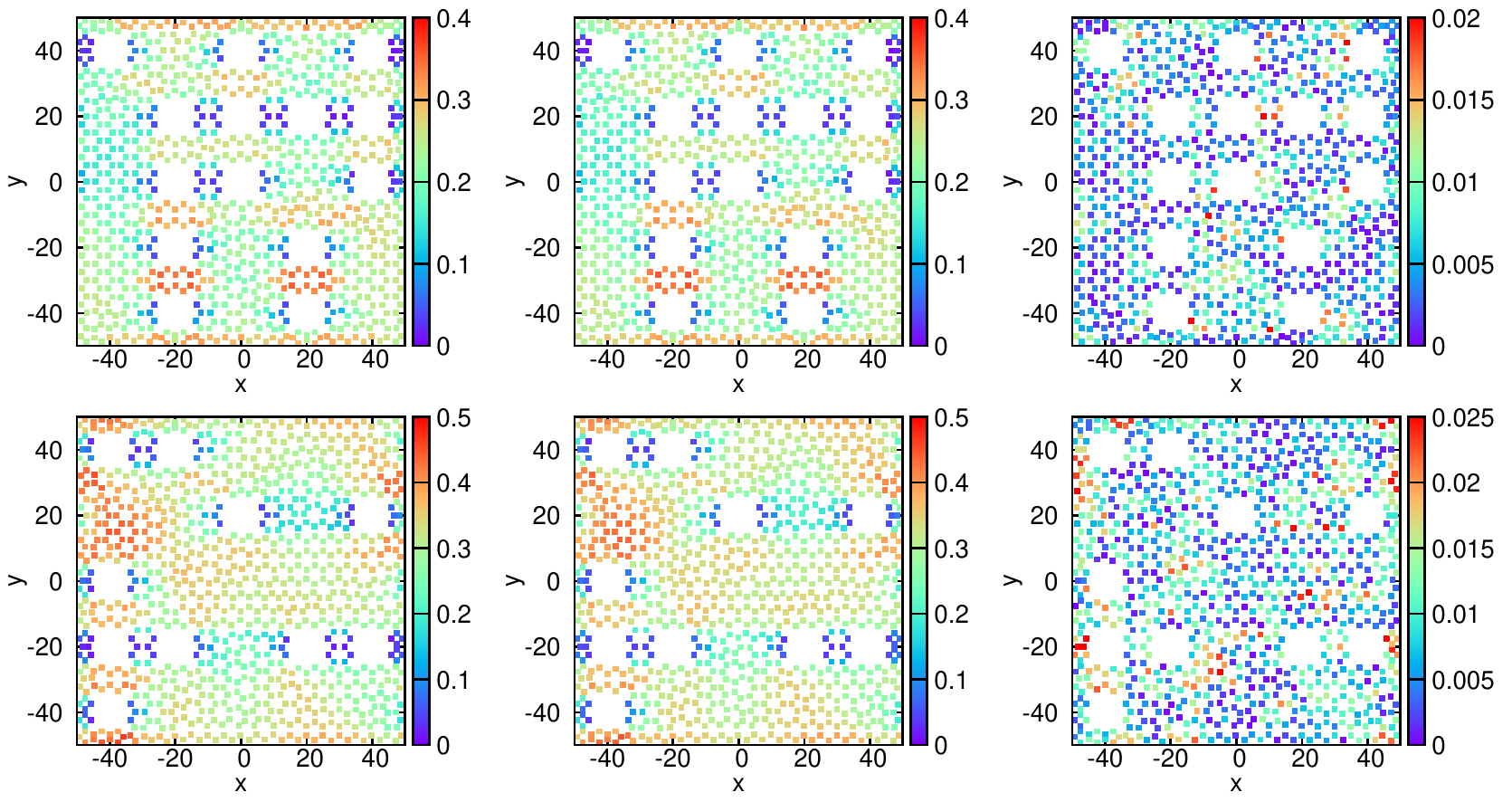}
    \caption{\textbf{Two examples of the predictions and pointwise absolute errors of multifidelity DeepONet for BTE.} The results in the same row is for the same geometry of pore locations. (Left) Reference solutions. (Middle) Predictions of multifidelity DeepONet. (Right) Pointwise absolute errors.}
    \label{fig:bte_mf_pred}
\end{figure}

\paragraph{Multifidelity DeepONet.} To train the high-fidelity DeepONet, we only use a dataset of size 1000. We use the same setup for DeepONet\textsubscript{H} as the DeepONet\textsubscript{L}, except that the branch net is chosen as a shallow network. Then the testing m.s.e. of the multifidelity DeepONet is $8.89 \pm 0.07 \times 10^{-5}$, and the $L^2$ relative error is 3.34$\pm$0.01\%. Furthermore, to verify Eq.~\eqref{eq:error}, we replace DeepONet\textsubscript{L} with the exact low-fidelity solver, then the testing m.s.e. of DeepONet\textsubscript{H} is $6.00 \pm 0.04 \times 10^{-5}$ and $L^2$ relative error is 2.72$\pm$0.01\%. Hence, we have
\begin{equation*}
    \underbrace{3.34 \pm 0.01\%}_{\mathcal{E}_{\mathcal{G}_H}} < \underbrace{1.72 \pm 0.02\%}_{\mathcal{E}_{\mathcal{G}_L}} + \underbrace{2.72 \pm 0.01\%}_{\mathcal{E}_{\mathcal{R}}}.
\end{equation*}
Two examples of the predictions and pointwise absolute errors of multifidelity DeepONet are shown in Fig.~\ref{fig:bte_mf_pred}.

\subsection{Boltzmann transport equation for inverse design}
\label{sec:BTE_inverse}

We have trained a multifidelity DeepONet as a surrogate model of BTE. Next we will use it for inverse design of thermal transport by optimizing an objective function $\mathcal{J}$:
$$\max_{v} \mathcal{J}(u).$$
Because $v_{i,j}$ is a binary value, this is a discrete optimization problem, and thus we can use algorithms like genetic algorithm. Here, we also consider to optimize the continuous relaxation via topology optimization. In this section, we consider several different objective functions, and we note that the same surrogate model allows for using different objective functions and constraints without retraining.

\subsubsection{Rationality of multifidelity DeepONet for continuous optimization algorithms}

In topology optimization, we treat $v_{ij}$ as a continuous variable defined on $[0, 1]$. Our surrogate model has the capability to predict for any value of $v_{i,j} \in \mathbb{R}$, but we only trained the network with $v_{i,j} \in \{0, 1\}$, and thus there is no guarantee that the trained network can predict reasonable values for $v_{i,j} \in (0, 1)$. Hence, in order to use it, we need to check whether the prediction of the trained multifidelity DeepONet for $v_{i,j} \in (0, 1)$ is reasonable.

\begin{figure}[htbp]
    \centering
    \includegraphics{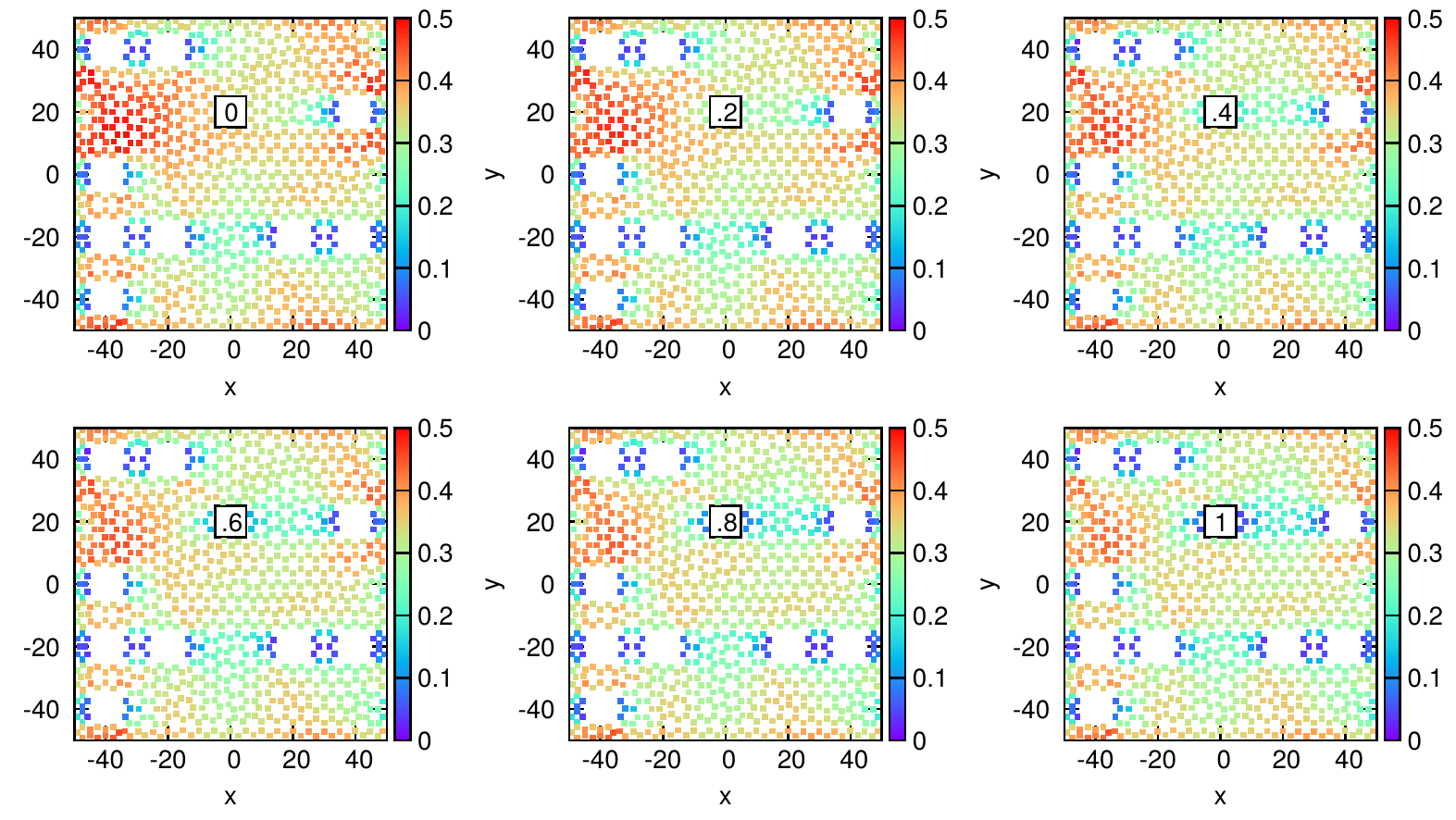}
    \caption{\textbf{Examples of predictions for $v_{2,3} \in [0, 1]$ by the multifidelity DeepONet.} $v_{2,3}$ is equal to 0 (not pore), 0.2, 0.4, 0.6, 0.8, and 1 (pore).}
    \label{fig:bte_semi_pore}
\end{figure}

As an example, we take a random geometry shown in Fig.~\ref{fig:bte_semi_pore}, and the value of $v_{2,3}$ changes from 0 to 1 gradually. When $v_{2,3}$ is closer to 1, the flux nearby becomes closer to 0. Hence, DeepONet makes reasonable predictions for intermediate value of $v_{ij}$. However, we note that in general there is no guarantee for the prediction, because we do not have training data in this domain.

\subsubsection{Maximizing normalized heat flux at a given location}
\label{sec:center_norm}

In the first inverse design problem, we consider to maximize the normalized heat flux at a given grid location. Here, we consider the center location, i.e.,
\begin{equation} \label{eq:g_1}
   \mathcal{J}(u) = \frac{u(0,0)}{\bar{u}},
\end{equation}
where
\begin{equation} \label{eq:u_mean}
    \bar{u} = \frac{1}{\text{Area}(\Omega)} \int_\Omega u(x,y) dxdy
\end{equation}
is the average value of $u$. To compute the integral in $\bar{u}$, we use the midpoint rule as
\begin{equation*}
    \bar{u} \approx \frac{1}{N^2} \sum_{i=1}^{N^2} u(x_i,y_i),
\end{equation*}
where $\{(x_i, y_i)\}_{i=1}^{N^2}$ are the mid-points of an equispaced mesh of $N$ by $N$, and we choose $N=100$. We note that we can also use other numerical integration methods such as Gaussian quadrature, which has a high convergence rate for smooth functions, but in this problem, Gaussian quadrature has a slow convergence, because we use ReLU activation and the network has a bad smoothness.

\begin{figure}[htbp]
    \centering
    \includegraphics{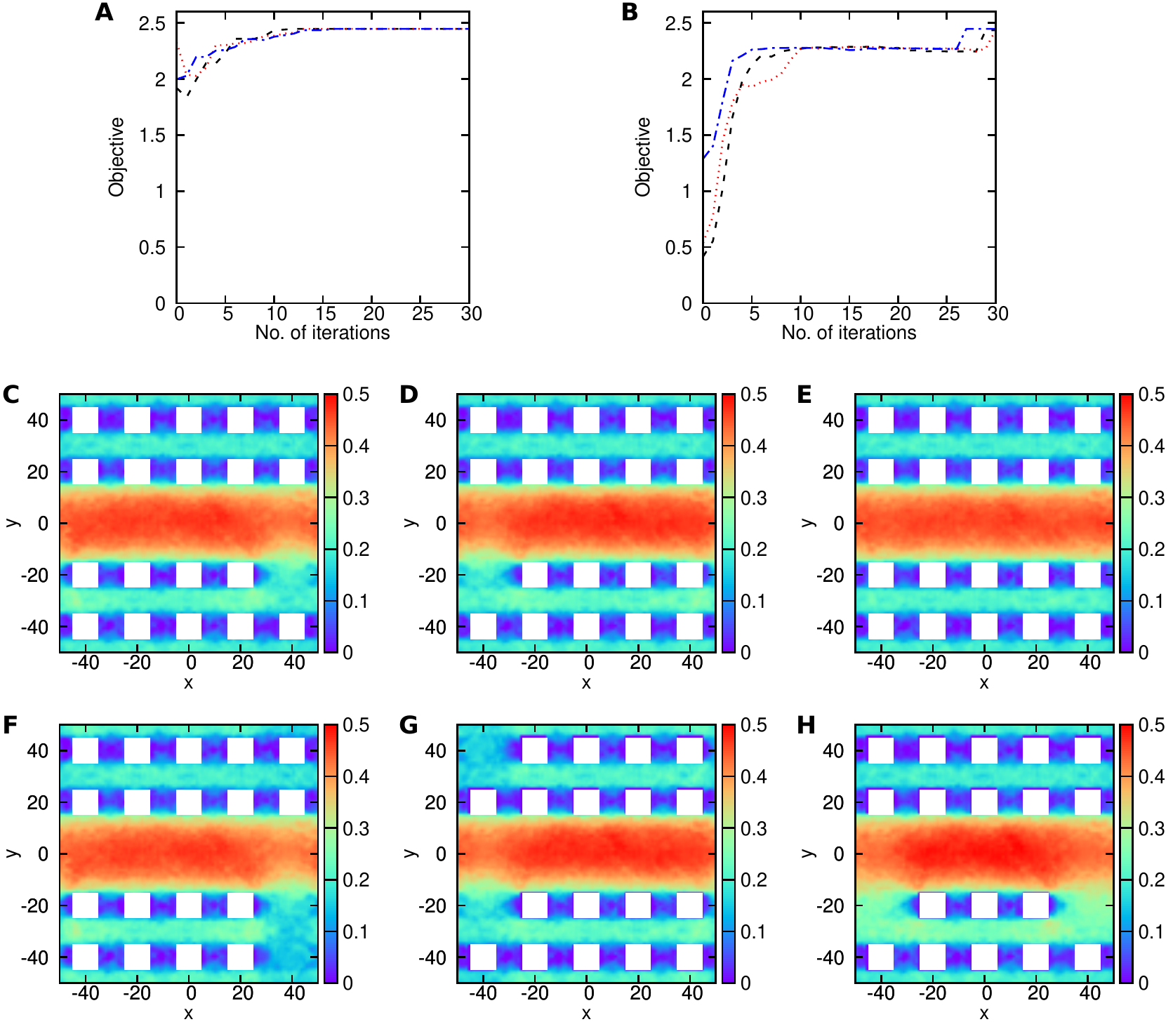}
    \caption{\textbf{Maximize the normalized heat flux at the center via GA and TO in Section~\ref{sec:center_norm}.} (\textbf{A}) The optimization process of GA. (\textbf{B}) The optimization process of TO. \textcolor{red}{Update} (\textbf{C}--\textbf{H}) The best 6 designs.}
    \label{fig:center_norm}
\end{figure}

We applied both GA and TO described in Section~\ref{sec:inverse} to solve this problem. Three examples of the values of the objective function during the iterative process of GA and TO are shown in Figs.~\ref{fig:center_norm}A and B, respectively, and they achieved similar final objective values of $\sim$2.4. The abrupt jumps in the optimization curves for TO correspond to the final evaluation of the fully binarized structure, which happens once the optimization has converged.

Because the learned multifidelity DeepONet has an error of 3.34\% as shown in Section~\ref{sec:bte_mf_deeponet}, in general the best design based on the multifidelity DeepONet may not be the best design according to the high-fidelity numerical solver. Hence, instead of only using the best design from GA or TO, we need to output several top candidates and then verify them using the numerical solver. Here, we consider the top 6 candidate designs (Figs.~\ref{fig:center_norm}C to H).

For each design in Fig.~\ref{fig:center_norm}, the objective values computed by the multifidelity DeepONet are listed in Table~\ref{tab:center_norm}. We also verify the objective values by using the high-fidelity numerical solver. Among these designs, the best one is the design in Fig.~\ref{fig:center_norm}E with the optimal objective $\mathcal{J}_\text{opt} \approx 2.343$. We also compute the relative difference of other designs compared to this best design $\frac{|\mathcal{J} - \mathcal{J}_{\text{opt}}|}{\mathcal{J}_{\text{opt}}}$. The best design (Fig.~\ref{fig:center_norm}E) is the third best based on the multifidelity DeepONet, but the best and second best designs (Figs.~\ref{fig:center_norm}C and D) are only 1.58\% and 2.18\% worse, respectively, which is consistent with the error of the multifidelity DeepONet.

\begin{table}[htbp]
    \centering
    \begin{tabular}{c|ccc}
    \toprule
    Design & Multifidelity DeepONet & Numerical solver & $\frac{|\mathcal{J} - \mathcal{J}_{\text{opt}}|}{\mathcal{J}_{\text{opt}}}$ \\
    \midrule
    Fig.~\ref{fig:center_norm}C & 2.447 & 2.306 & 1.58\% \\
    Fig.~\ref{fig:center_norm}D & 2.440 & 2.292 & 2.18\% \\
    Fig.~\ref{fig:center_norm}E & 2.435 & \textbf{2.343} & Optimal \\
    Fig.~\ref{fig:center_norm}F & 2.425 & 2.233 & 4.69\% \\
    Fig.~\ref{fig:center_norm}G & 2.414 & 2.235 & 4.61\% \\
    Fig.~\ref{fig:center_norm}H & 2.412 & 2.211 & 5.63\% \\
    \bottomrule
    \end{tabular}
    \caption{\textbf{Maximize the normalized heat flux at the center via GA and TO in Section~\ref{sec:center_norm}.} The objective values of the found designs (Fig.~\ref{fig:center_norm}) computed by the multifidelity DeepONet and the numerical solver.}
    \label{tab:center_norm}
\end{table}

\subsubsection{Maximizing normalized heat flux on multiple points}
\label{sec:path}

\begin{figure}[htbp]
    \centering
    \includegraphics{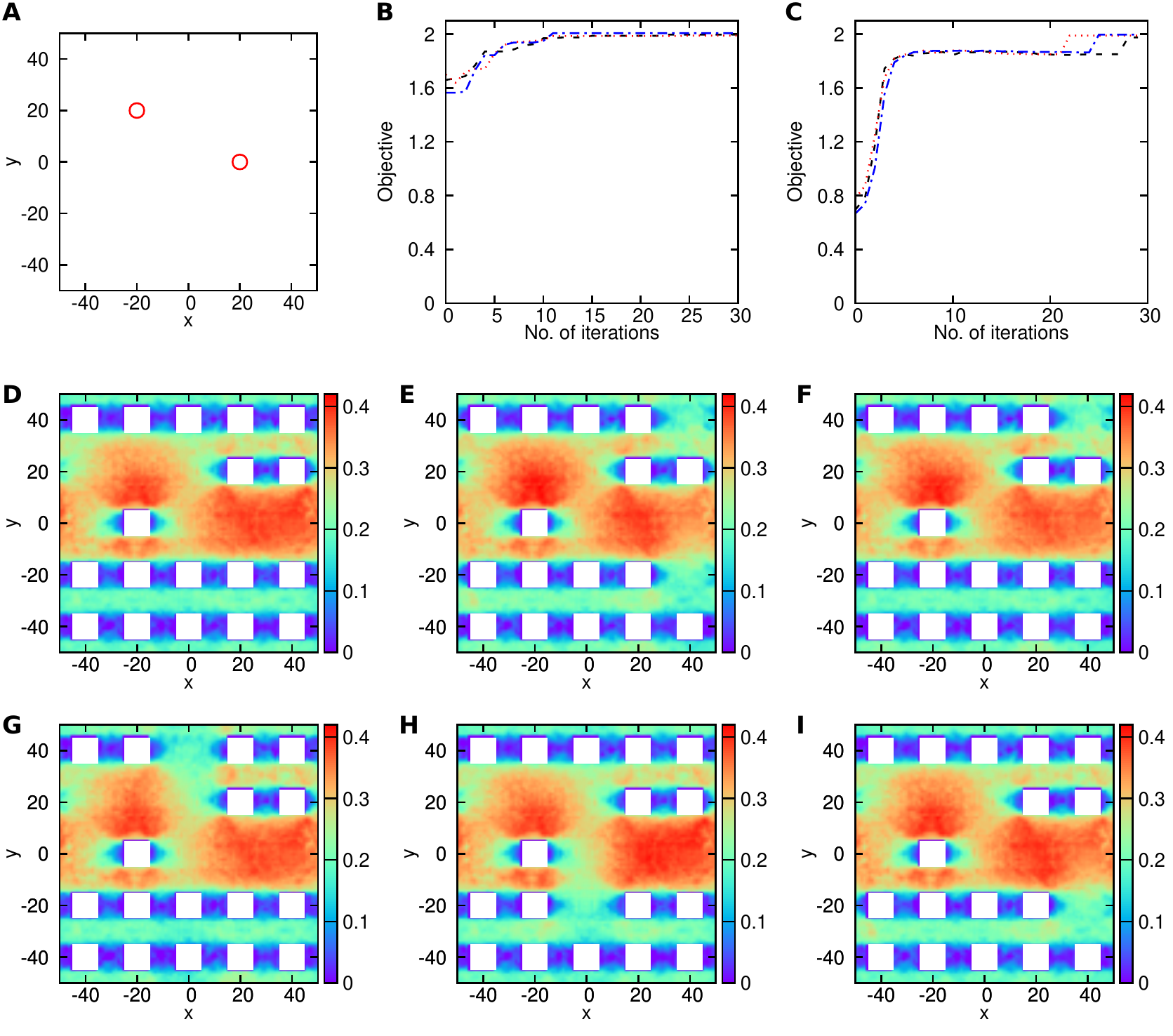}
    \caption{\textbf{Maximize the normalized heat flux on multiple points in Section~\ref{sec:path}.} (\textbf{A}) The locations of the points to be optimized. (\textbf{B}) The optimization process of GA. (\textbf{C}) The optimization process of TO. \textcolor{red}{Update} (\textbf{D}--\textbf{I}) The best 6 designs.}
    \label{fig:path}
\end{figure}

We have considered the objective for a single location, and here we consider the objective for multiple points. Specifically, we consider 2 point locations: $\mathbf{x}_1 = (-20, 20)$ and $\mathbf{x}_2=(20, 0)$ (the red circles in Fig.~\ref{fig:path}A), and the objective function is
$$\mathcal{J} = \frac{1}{2} \sum_{i=1}^2 \frac{u(\mathbf{x}_i)}{\bar{u}},$$
where $\bar{u}$ is defined in Eq.~\eqref{eq:u_mean}.

GA and TO achieved similar objective values of $\sim$2 (Figs.~\ref{fig:path}B and C), and we show the top 6 candidate designs (Figs.~\ref{fig:path}D to I). For each design, the objective values computed by the multifidelity DeepONet and the numerical solver are listed in Table~\ref{tab:path}. Among these designs, the best design (Fig.~\ref{fig:path}D) found by the multifidelity DeepONet is indeed the best with the optimal objective $\mathcal{J}_\text{opt} \approx 1.987$. We also compute the relative difference of other designs compared to this design, and most of the relative differences are within 3\% (Table~\ref{tab:path}).

\begin{table}[htbp]
    \centering
    \begin{tabular}{c|ccc}
    \toprule
    Design & Multifidelity DeepONet & Numerical solver & $\frac{|\mathcal{J} - \mathcal{J}_{\text{opt}}|}{\mathcal{J}_{\text{opt}}}$ \\
    \midrule
    Fig.~\ref{fig:path}D & 2.008 & \textbf{1.987} & Optimal \\
    Fig.~\ref{fig:path}E & 1.996 & 1.928 & 2.97\% \\
    Fig.~\ref{fig:path}F & 1.993 & 1.941 & 2.32\% \\
    Fig.~\ref{fig:path}G & 1.990 & 1.961 & 1.31\% \\
    Fig.~\ref{fig:path}H & 1.990 & 1.951 & 1.81\% \\
    Fig.~\ref{fig:path}I & 1.990 & 1.964 & 1.16\% \\
    \bottomrule
    \end{tabular}
    \caption{\textbf{Maximize the normalized heat flux on multiple points in Section~\ref{sec:path}.} The objective values of the designs (Fig.~\ref{fig:path}) computed by the multifidelity DeepONet and the numerical solver.}
    \label{tab:path}
\end{table}

\subsubsection{Maximizing normalized heat flux on multiple points with a constraint on the number of pores}
\label{sec:path_constraint}

In Section~\ref{sec:path}, the best design is the one in Fig.~\ref{fig:path}D, and the results show that if we add additional pores, then the objective becomes smaller. Here, we aim to investigate the best design given an extra constraint on the number of pores. Specifically, we require that
$$\text{No.\ of pores} \leq N$$
for a given integer $N$.

To consider this constraint, for GA, we simply return 0 as the objective if the design does not satisfy the constraint. For TO, the implementation is less straighforward, because the objective function needs to stay differentiable. We discuss how to implement the constraint in TO as follows. We implemented the constraint by adding a differentiable non-linear constraint on the sum of the projected density. Note that since the constraint is on the projected function using Eq.~\eqref{eq:thresholding}, the density is not fully binarized. In practice, we constrained the sum to be less than $N-0.5$, which resulted in designs with less than $N$ pores most of the time.

Here, we consider three cases ($N=11$, $N=9$, and $N=7$) and follow the same procedure as in Section~\ref{sec:path} to find the best design for each case. For $N=11$, the best 6 designs and the corresponding objective values are in Fig.~\ref{fig:path_constraint_N11} and Table~\ref{tab:path_constraint_N11}, respectively. For $N=9$, the best 6 designs and the corresponding objective values are in Fig.~\ref{fig:path_constraint_N9} and Table~\ref{tab:path_constraint_N9}, respectively. For $N=7$, the best 6 designs and the corresponding objective values are in Fig.~\ref{fig:path_constraint_N7} and Table~\ref{tab:path_constraint_N7}, respectively.

\begin{figure}[htbp]
    \centering
    \includegraphics{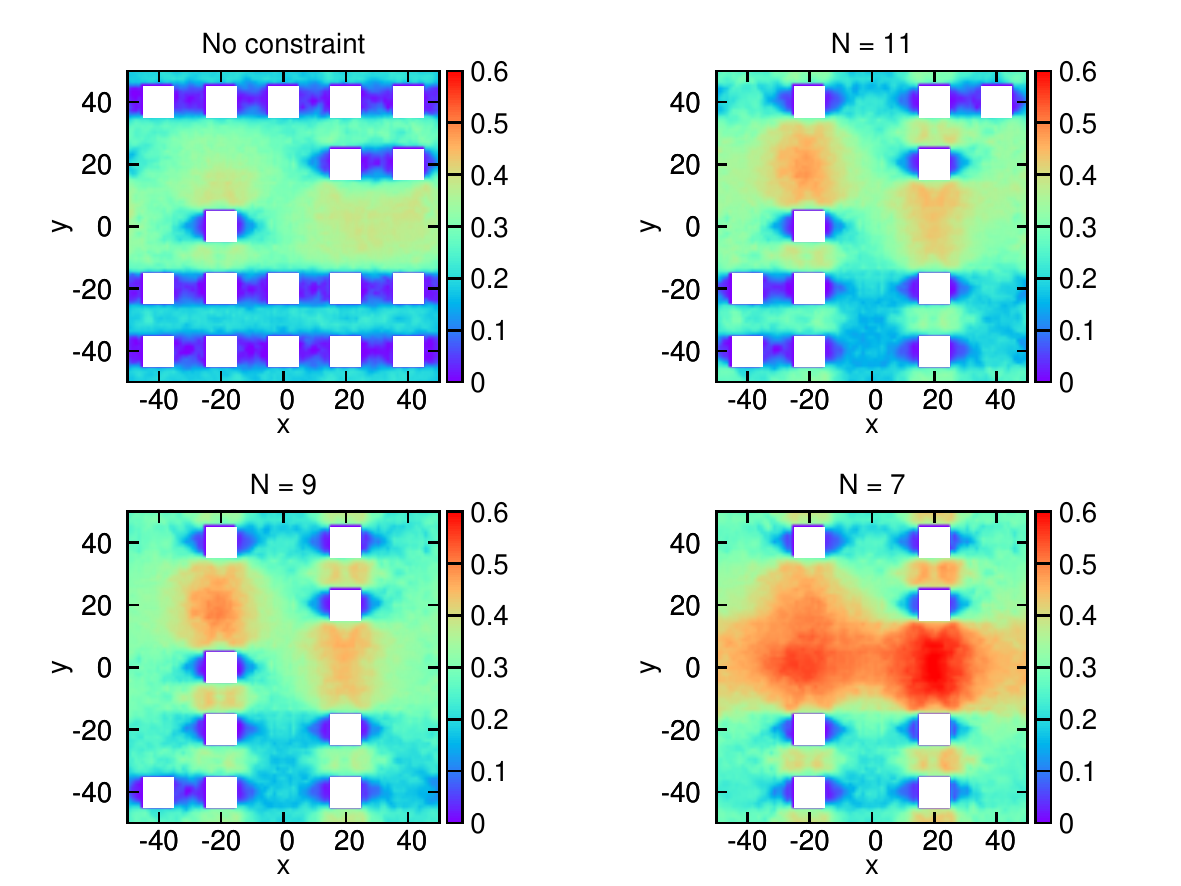}
    \caption{\textbf{Maximize the normalized heat flux on multiple points with a constraint on the number of pores in Section~\ref{sec:path_constraint}.} The best designs for different values of $N$.}
    \label{fig:path_constraint}
\end{figure}

The best designs for all the cases are listed in Fig.~\ref{fig:path_constraint}. It is interesting that the pore locations of the best design for $N=11$ is a subset of the pore locations for the case without any constraint. This is also true for the pair of $N=11$ and $N=9$, and the pair of $N=9$ and $N=7$. For the value of the objective function, a stronger constraint on the number of pores makes the objective value smaller (Table~\ref{tab:path_constraint}).

\begin{table}[htbp]
    \centering
    \begin{tabular}{c|ccc}
    \toprule
    $N$ & Multifidelity DeepONet & Numerical solver \\
    \midrule
    No constraint & 2.008 & 1.987 \\
    \midrule
    11 & 1.879 & 1.857 \\
    9 & 1.820 & 1.763 \\
    7 & 1.678 & 1.639 \\
    \bottomrule
    \end{tabular}
    \caption{\textbf{Maximize the normalized heat flux on multiple points with a constraint on the number of pores in Section~\ref{sec:path_constraint}.} The objective values of the designs in Fig.~\ref{fig:path_constraint} computed by the multifidelity DeepONet and the numerical solver.}
    \label{tab:path_constraint}
\end{table}

\section{Conclusion}
\label{sec:conclusion}

In this study, we developed a new version of DeepONet for learning efficiently from multifidelity datasets. A multifidelity DeepONet includes two independent DeepONets coupled by residual learning and input augmentation. We demonstrated that, compared to single-fidelity DeepONet, multifidelity DeepONet achieves one order of magnitude smaller error when using the same amount of high-fidelity data. By combining multifidelity DeepONet with genetic algorithms or topology optimization, we can reuse the trained network for fast inverse design of multiple objective functions.

We have introduced two approaches of input augmentation, where Approach~I uses the entire function of $\mathcal{G}_L(v)$, and Approach~II uses only the low-fidelity prediction at one point $\mathcal{G}_L(v)(\xi)$. In this study, we find that Approach~II is more accurate than Approach~I, but there are also other possibilities that might be fruitful to explore: for example, using the low-fidelity solution in a local domain $[\xi - \Delta \xi, \xi + \Delta \xi]$. Moreover, the two sub-DeepONets used in the multifidelity DeepONet are ``vanilla'' DeepONets. Several extensions of DeepONet have been developed very recently, as discussed in the introduction, and these extensions may further improve the performance of multifidelity DeepONet.

% Other approaches such as transfer learning will be investigate in the future.

\section*{Acknowledgments}

This work was partially supported by MIT-IBM Watson AI Laboratory (Challenge No.~2415). RP is also
supported in part by the U.S. Army Research Office through the Institute for Soldier Nanotechnologies at MIT under Award Nos.~W911NF-18-2-0048 and W911NF-13-D-0001.

\appendix

\section{Best designs in Section~\ref{sec:path_constraint}}

\begin{figure}[htbp]
    \centering
    \includegraphics{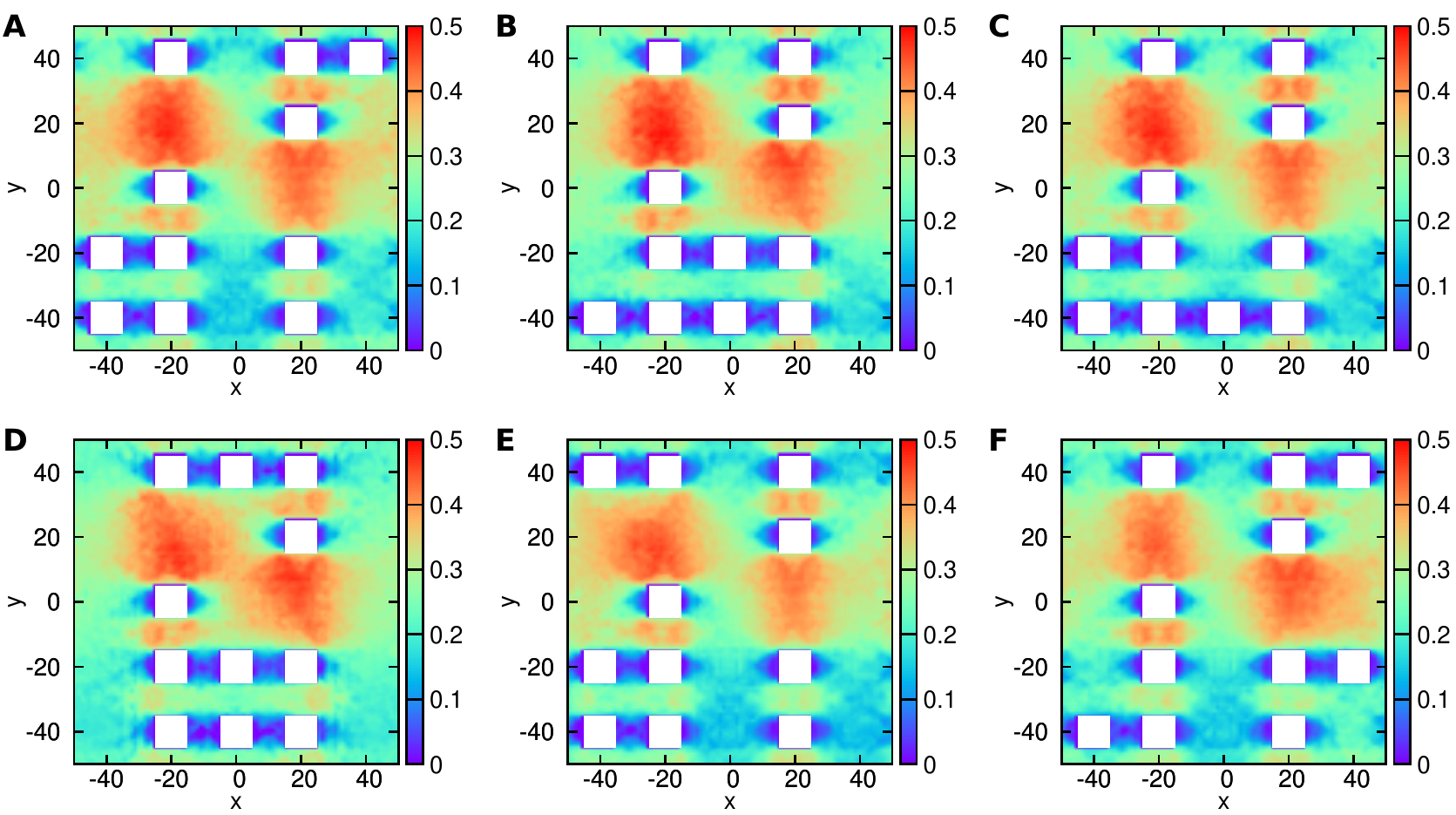}
    \caption{\textbf{Maximize the normalized heat flux on multiple points with a constraint on the number of pores in Section~\ref{sec:path_constraint}.} The best 6 designs for $N=11$.}
    \label{fig:path_constraint_N11}
\end{figure}

\begin{table}[htbp]
    \centering
    \begin{tabular}{c|ccc}
    \toprule
    Design & Multifidelity DeepONet & Numerical solver & $\frac{|\mathcal{J} - \mathcal{J}_{\text{opt}}|}{\mathcal{J}_{\text{opt}}}$ \\
    \midrule
    Fig.~\ref{fig:path_constraint_N11}A & 1.879 & \textbf{1.857} & Optimal \\
    Fig.~\ref{fig:path_constraint_N11}B & 1.879 & 1.815 & 2.26\% \\
    Fig.~\ref{fig:path_constraint_N11}C & 1.873 & 1.840 & 0.92\% \\
    Fig.~\ref{fig:path_constraint_N11}D & 1.872 & 1.815 & 2.26\% \\
    Fig.~\ref{fig:path_constraint_N11}E & 1.869 & 1.826 & 1.67\% \\
    Fig.~\ref{fig:path_constraint_N11}F & 1.866 & 1.826 & 1.67\% \\
    \bottomrule
    \end{tabular}
    \caption{\textbf{Maximize the normalized heat flux on multiple points with a constraint on the number of pores in Section~\ref{sec:path_constraint}.} The objective values of the designs (Fig.~\ref{fig:path_constraint_N11}) computed by the multifidelity DeepONet and the numerical solver.}
    \label{tab:path_constraint_N11}
\end{table}

\begin{figure}[htbp]
    \centering
    \includegraphics{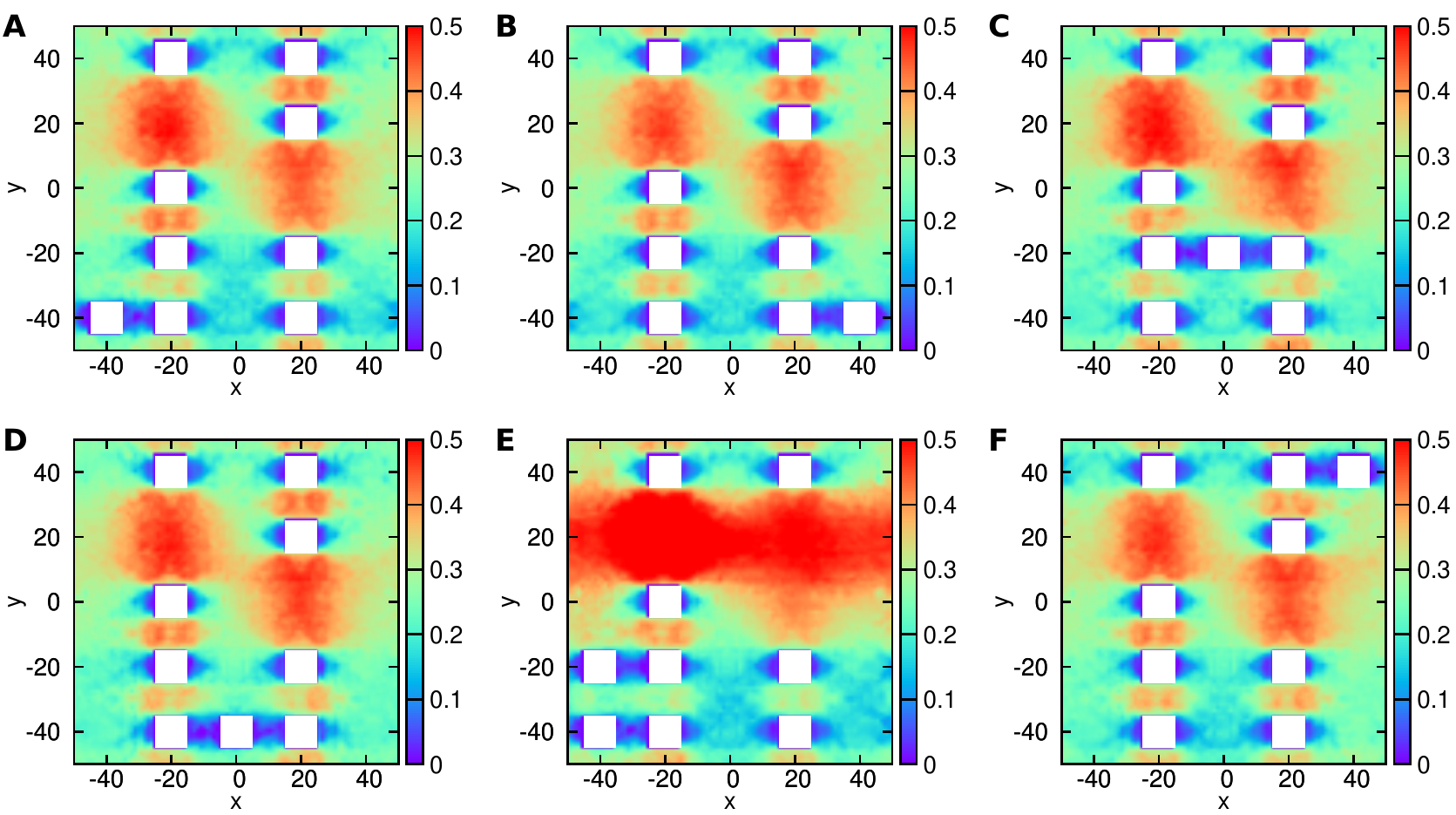}
    \caption{\textbf{Maximize the normalized heat flux on multiple points with a constraint on the number of pores in Section~\ref{sec:path_constraint}.} The best 6 designs for $N=9$.}
    \label{fig:path_constraint_N9}
\end{figure}

\begin{table}[htbp]
    \centering
    \begin{tabular}{c|ccc}
    \toprule
    Design & Multifidelity DeepONet & Numerical solver & $\frac{|\mathcal{J} - \mathcal{J}_{\text{opt}}|}{\mathcal{J}_{\text{opt}}}$ \\
    \midrule
    Fig.~\ref{fig:path_constraint_N9}A & 1.820 & \textbf{1.763} & Optimal \\
    Fig.~\ref{fig:path_constraint_N9}B & 1.810 & 1.760 & 0.17\% \\
    Fig.~\ref{fig:path_constraint_N9}C & 1.798 & 1.747 & 0.91\% \\
    Fig.~\ref{fig:path_constraint_N9}D & 1.798 & 1.759 & 0.23\% \\
    Fig.~\ref{fig:path_constraint_N9}E & 1.797 & 1.711 & 2.95\% \\
    Fig.~\ref{fig:path_constraint_N9}F & 1.788 & 1.758 & 0.28\% \\
    \bottomrule
    \end{tabular}
    \caption{\textbf{Maximize the normalized heat flux on multiple points with a constraint on the number of pores in Section~\ref{sec:path_constraint}.} The objective values of the designs (Fig.~\ref{fig:path_constraint_N9}) computed by the multifidelity DeepONet and the numerical solver.}
    \label{tab:path_constraint_N9}
\end{table}

\begin{figure}[htbp]
    \centering
    \includegraphics{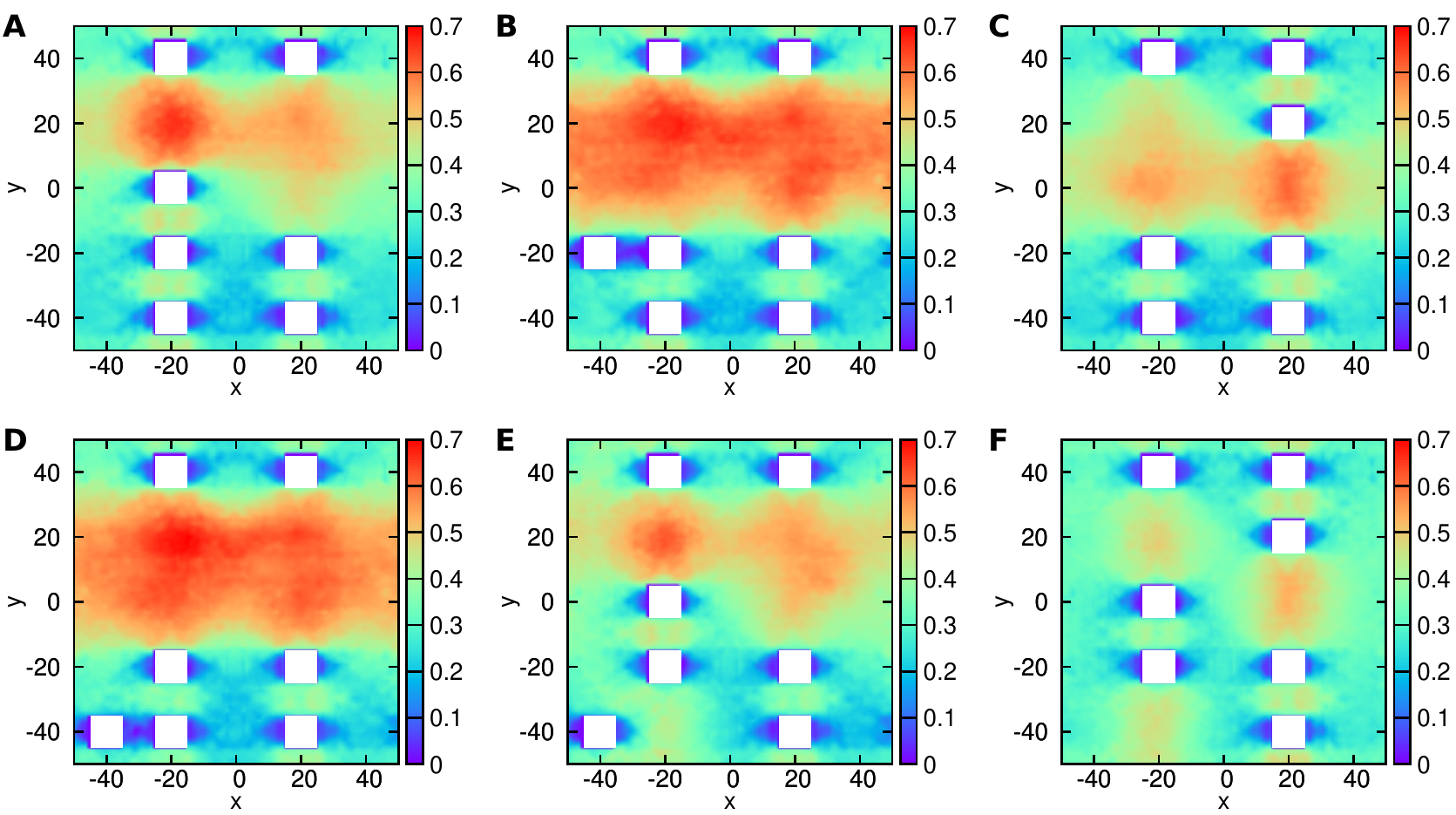}
    \caption{\textbf{Maximize the normalized heat flux on multiple points with a constraint on the number of pores in Section~\ref{sec:path_constraint}.} The best 6 designs for $N=7$.}
    \label{fig:path_constraint_N7}
\end{figure}

\begin{table}[htbp]
    \centering
    \begin{tabular}{c|ccc}
    \toprule
    Design & Multifidelity DeepONet & Numerical solver & $\frac{|\mathcal{J} - \mathcal{J}_{\text{opt}}|}{\mathcal{J}_{\text{opt}}}$ \\
    \midrule
    Fig.~\ref{fig:path_constraint_N7}A & 1.708 & 1.633 & 0.366\% \\
    Fig.~\ref{fig:path_constraint_N7}B & 1.680 & 1.572 & 4.087\% \\
    Fig.~\ref{fig:path_constraint_N7}C & 1.678 & \textbf{1.639} & Optimal \\
    Fig.~\ref{fig:path_constraint_N7}D & 1.677 & 1.568 & 4.331\% \\
    Fig.~\ref{fig:path_constraint_N7}E & 1.677 & 1.556 & 5.064\% \\
    Fig.~\ref{fig:path_constraint_N7}F & 1.666 & 1.572 & 4.087\% \\
    \bottomrule
    \end{tabular}
    \caption{\textbf{Maximize the normalized heat flux on multiple points with a constraint on the number of pores in Section~\ref{sec:path_constraint}.} The objective values of the designs (Fig.~\ref{fig:path_constraint_N7}) computed by the multifidelity DeepONet and the numerical solver.}
    \label{tab:path_constraint_N7}
\end{table}

\bibliographystyle{unsrt}
\bibliography{main}

\end{document}